\documentclass[preprint,12pt,3p]{elsarticle}
\usepackage{graphicx}
\usepackage{graphics}
\usepackage[table]{xcolor}
\usepackage{array}
\usepackage{enumitem}
\usepackage{amsmath}
\usepackage[utf8]{inputenc}
\usepackage[T1]{fontenc}
\usepackage{afterpage}
\usepackage{amssymb}
\usepackage{setspace}
\usepackage[left]{lineno}
\newcolumntype{+}{!{\vrule width 2pt}}
\usepackage{xcolor}
\usepackage{color,soul}
\biboptions{sort&compress}
\usepackage{eurosym}
\usepackage{hyperref}
\usepackage{gensymb}
\hypersetup{
    colorlinks=False,
    linkcolor=blue,
    filecolor=magenta,      
    urlcolor=cyan}
\DeclareUnicodeCharacter{20AC}{\EUR}
\usepackage{xparse}
\usepackage{amsfonts}
\usepackage{booktabs}
\usepackage{siunitx}

\journal{Biomaterials}
\begin{document}
%\doublespacing
%\linenumbers
\begin{frontmatter}
\clearpage

\title{Decoding the Rejuvenating Effects of Mechanical Loading on Skeletal Maturation using in Vivo Imaging and Deep Learning}

\author[label1,label2]{Pouyan Asgharzadeh}
\address[label1]{Institute for Modelling and Simulation of Biomechanical Systems, University of Stuttgart, Stuttgart, Germany}
\address[label2]{Stuttgart Center for Simulation Science (SC SimTech), Stuttgart, Germany}
\author[label1,label2]{Oliver Röhrle \corref{cor1}}
\ead{roehrle@simtech.uni-stuttgart.de}
\ead[url]{https://www.mechbau.uni-stuttgart.de/ls2/jrg/}
\author[label3]{Bettina M. Willie}
\address[label3]{Research Centre, Shriners Hospital for Children-Canada, Department of Pediatric Surgery, McGill University, Canada}
\author[label1,label2]{Annette I. Birkhold \fnref{t1}}

\cortext[cor1]{Corresponding author. Tel.: 0049 711 685-66284
  Fax: 0049 711 685-66347 }
\fntext[t1]{Currently an employee of Siemens Healthcare GmbH}

\thispagestyle{empty}
\begin{abstract} 
Throughout the process of aging, deterioration of bone macro- and micro-architecture, as well as material decomposition result in a loss of strength and therefore in an increased likelihood of fractures. To date, precise contributions of age-related changes in bone (re)modeling and (de)mineralization dynamics and its effect on the loss of functional integrity are not completely understood. Here,  we  present an image-based deep learning approach to quantitatively describe the dynamic effects of short-term aging and adaptive response to treatment in proximal mouse tibia and fibula. Our approach allowed us to perform an end-to-end age prediction based on $\mu$CT images to determine the dynamic biological process of tissue maturation during a two week period, therefore permitting a short-term bone aging prediction with $95\%$ accuracy. In a second application, our radiomics analysis reveals  that two weeks of in vivo mechanical loading are associated  with an underlying rejuvenating effect of 5 days. Additionally, by quantitatively analyzing the learning process, we could, for the first time, identify the localization of the age-relevant encoded information and demonstrate $89\%$ load-induced similarity of these locations in the loaded tibia with younger bones. These  data  suggest  that our method enables identifying  a  general prognostic phenotype of a certain bone age as well as a temporal and localized loading-treatment effect on this apparent bone age.  Future translational applications of this method may provide  an  improved decision-support method for osteoporosis treatment at low cost.

\end{abstract}

\end{frontmatter}

\thispagestyle{empty}

%\doublespacing
\newpage
\section{Introduction}
\label{intro}

Bone is a hierarchical, dynamic, living tissue whose primary function is mechanical integrity, providing protection to internal organs and enabling mobility. Internal and external stimuli cause continuous (re)modeling making bone a highly dynamic structure. Both, bone stiffness and strength depend on properties such as mass and shape, the distribution of mass (geometry and architecture), micro-architecture and microscopic material property distribution \cite{Seeman2006}. Human and animal studies show that skeletal maturation and aging affect both, bone micro-architecture \cite{szulc2006bone,riggs2008population,Willie2013} and tissue material properties \cite{Roschger2008,Koehne2014}. Formation and resorption dynamics in trabecular \cite{Birkhold2014} and cortical bone \cite{Birkhold2014b} are  altered with aging in a site-specific manner \cite{birkhold2017tomography,birkhold2016periosteal}. 
As a result, with increasing age a net bone loss occurs \cite{szulc2006bone,ahlborg2003bone}, often resulting in osteoporosis and a subsequent increase in fragility fracture risk \cite{Kanis2002}. The rules governing age-related alterations in bone composition, organization, and elasticity across structural hierarchies are, however, to date not completely understood.
%Disease, like osteoporosis, or age  change the dynamical processes of material decomposition in bones that lead to changes in the micro- and macro-structure, in bone strength and, subsequently to an increase in the likelihood of fracture \cite{demontiero2012aging}.

Given the fact that osteoporosis causes worldwide more than 8.9 million fractures per year \cite{hernlund2013osteoporosis}, it is essential to develop a precise and comprehensive analysis of phenotypic changes and abnormalities at all relevant length scales. Assessing the onset of osteoporosis and disease progression is therefore challenging. Within clinical practice, dual energy X-ray absorptiometry (DXA) and biochemical markers remain the standard methods of monitoring osteoporotic patients receiving pharmacological treatments. The T-score is derived from measurements of the areal bone mineral density (aBMD), which is obtained by DXA \cite{kanis2008european}. DXA is a useful clinical tool, but has several limitations including restriction to a two-dimensional image, lack of distinction between trabecular and cortical bone, lack of information on bone microarchitecture, difficulties in edge detection and projection artefacts. Additionally, the predictive ability of this method is low \cite{Riggs2002,Bolotin2007} with less than half of all nonvertebral fractures occurring in postmenopausal women having an osteoporotic T-score \cite{schuit2004fracture}.  Biochemical markers are indirect indicators of the rates of formation and resorption of bone and give no insight into its quality or mechanical properties. Furthermore, like all biochemical markers they are subject to pre-analytical, analytical and post-analytical sources of variability and the results may be affected by a range of non-skeletal conditions. 
High-resolution peripheral quantitative computed tomography (HR-pQCT) is emerging as a powerful non-invasive bone imaging modality capable of assessing volumetric BMD, microarchitecture and strength, and distinguishing cancellous and cortical bone. Additionally, micro-finite element and homogenized finite element models based on HR-pQCT imaging are increasingly used to predict bone stiffness and strength \cite{dall2012qct,zysset2013finite}. The Bone Microarchitecture International Consortium (BoMIC), which combined individual-level prospective data from eight cohorts (7254 individuals, mean age: $69\pm9$ years), recently reported that HR-pQCT parameters improved fracture prediction beyond femoral neck aBMD or fracture risk assessment tool (FRAX) scores alone \cite{samelson2019cortical}.

Preclinical studies using micro-computed tomography ($\mu$CT) aiming to assess bone maturation and (re)modeling have focused on selectively extracting mechanical or morphological  features such as mineralization \cite{ferguson2003bone} or bone volume \cite{halloran2002changes,ko2011deterioration,moustafa2009mouse} and their alterations \cite{lukas2013mineralization,birkhold2017tomography}. Although these approaches decode certain aspects of structural changes in bone, they neglect the underlying interplay and concurrency of (re)modeling and (de)mineralization. The measures extracted from these properties are selective and therefore not sufficient to predict fracture in diseases such as osteoporosis. To provide more precise descriptions of the disease phenotype, the diverse manifestations must be captured allowing one to distinguish healthy bones from diseased ones and young bones from aged ones to define disease onset and progression into sub-classes. This would permit a much more precise understanding of bone quality, as well as a better prediction of fracture risk and treatment outcome.

% Using machine learning for bone assessment
A major challenge in disease diagnosis is interpreting information-rich (imaging) data.
This challenge is at the same time a great opportunity, as there exits nowadays artificial intelligence-based methods that have the capabilities and power to analyze relationships within rich datasets, e.g., relationships of particular dynamic biological phenomena. Artificial intelligence, for example, has been used to diagnose Alzheimer disease based on Magnetic Resonance Imaging (MRI) \cite{suk2014hierarchical} or to analyze skin lesion for diagnosing malignancy \cite{esteva2017dermatologist}. 
Similar to the previous two examples, one can also use artificial intelligence to analyze (re)modeling of bone using X-ray images (eg. $\mu$CT, HR-pQCT).
As scatter and attenuation information of $\mu$CT images contain information about material composition, distribution and amount, they potentially contain all structural information that is needed to asses bone maturation \cite{mosekilde1988age}. 
Despite the fact that recent studies can extract from 2D and 3D X-ray image data more features describing bone quality through assessment of vBMD and microstructure ~\cite{burghardt2010reproducibility,bouxsein2010guidelines,mader2013quantitative,birkhold2015monitoring,ruegsegger1976quantification,macdonald2011age}, information on bone (re)modelling rates are only obtained through invasive histomophometry analysis of iliac crest bone biopsies. Fortunately, recent advances in artificial intelligence towards deep learning now enable further data analysis by utilizing high-throughput image data.
Compared to traditional machine learning methods~\cite{shen2017deep,litjens2017survey}, deep learning methods do not only exhibit an improved prediction accuracy, but also provide the ability to visualize learned features, to link discovered features with clinical relevance. 
The first applications for bone age assessment in pediatrics using deep neural networks (DNN) showed already some success in classifying/predicting bone age from 2D X-ray images \cite{spampinato2017deep,torres2017bone,lee2017fully}.
Further, they provide confidence that DNN-based methods can also provide insights into the underlying processes of skeletal maturation and bone (re)modeling.

% what we do / summary
In this study, we present a deep learning approach applied to 2D projection X-ray images of bones as an end-to-end tool for site-specific, spatio-temporal assessment of bone tissue maturation and intervention effects. By simultaneous evaluating several relevant hierarchies, our method allows us to reconstruct continuous biological processes such as aging or adaptation of bone. We developed and evaluated our method on pre-clinical $\mu$CT data of mouse bones and investigated bone adaptation in response to in vivo tibial compressive loading.
To do so, we first evaluate, if our method allows identifying short-term, skeletal maturation-related changes in the proximal tibiae and fibulae based on $\mu$CT images. 
Therefore, we analyze short-term (15 days) dynamic skeletal maturation processes in adult female mice bones. 
Second, we evaluate, if our model can be used to identify treatment results and relate these to load-induced surface (re)modeling ("rejuvenation")-effects.
Furthermore, by analyzing the learning process of our DNN through saliency maps, we quantify the spatial localization of the network attention, permitting determination of where in the bone the "skeletal age" information is manifested.

\section{Materials and Methods}
\subsection{In vivo Mechanical Loading}
\begin{figure}[htbp]
\centering
\includegraphics[width=\textwidth]{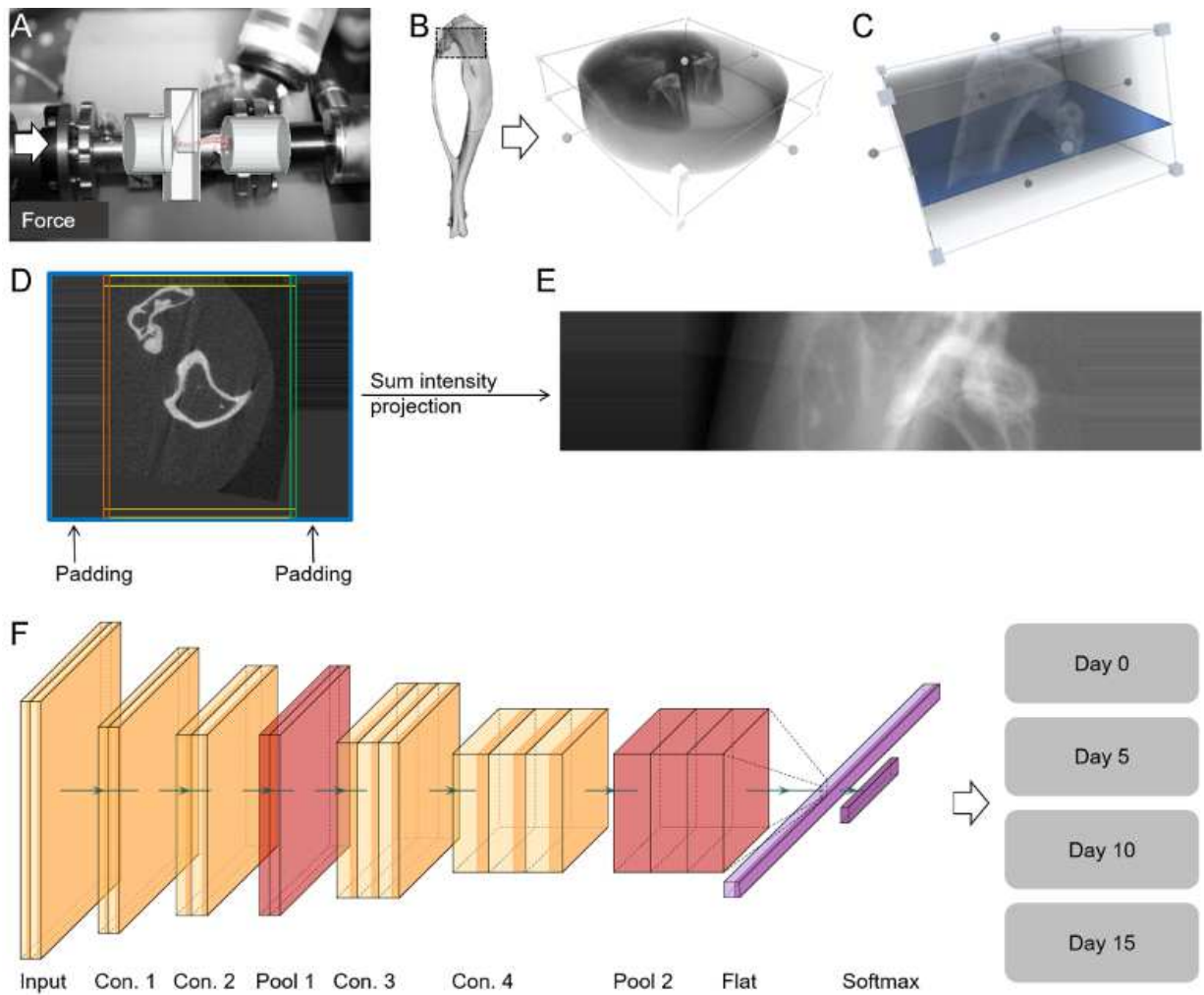}
\caption{Overall work flow and network architecture. (A) Experimental setup of applying load to the bone. (B) Imaged ROI and 3D raw data containing both right (control) and left (loaded) tibiae and fibulae. (C) Cropped 3D image containing only one tibia and fibula. Blue plane indicates an exemplary slice on which the padding step is applied. (D) Each slice (inside the colored stripes) is padded with border stripes of pixels (colored stripes: padding) in each direction: yellow and orange $\rightarrow$ posterior-anterior, red and green $\rightarrow$ lateral-medial directions. (E) Sum intensity projection of the 3D image in C after padding. (F) BAAM network architecture including convolutional layers (Con.), pooling layers (Pool), flattened layer (Flat), softmax layer and the four output classes.}
\label{fig:1}
\end{figure}

As reported in \cite{Birkhold2014,Willie2013}, $20$ mice (female C57Bl/6J, 26 weeks old at beginning of experiments, Jackson Laboratories, Sulzfeld, Germany) underwent two weeks of in vivo cyclic compressive loading of the left tibia. Loading was applied $5$ days/week (M-F) for $2$ weeks while mice were anesthetized ($216$ cycles applied daily at $4$Hz, delivering $-11$N loads, $1200\,\mu \epsilon$ on the medial surface of the tibial mid-shaft, determined by prior in vivo strain gauging experiments; Fig. \ref{fig:1}A). The right tibia served as physiologically loaded internal control. The first loading session occurred on the first day of in vivo imaging.

\subsection{In vivo Monitoring of Tissue Maturation and Adaptation}
\label{Materials}
In total, $79$ $\mu$CT image sets of both proximal tibiae and fibulae were collected during two weeks of tissue maturation (right limb) and adaptation (left limb). Both limbs, including the tibia and fibula, were scanned and combined within one imaging procedure (cf.~Fig.~\ref{fig:1}B). 
In vivo $\mu$CT was performed at an isotropic voxel size of $10.5\,\mu m$ (vivaCT40, Scanco Medical, Switzerland; $55$ kVp, $145$ mA, $600\,ms$ integration time, no frame averaging). 
The mice were scanned starting from the growth plate and was
extended for $432$ slices ($4536\,\mu m$) in the distal direction. To prevent motion artifacts and obtain reproducible scan regions and bone orientations, mice were anesthetized and kept during the scans in a fixed position using a custom-made mouse bed. In time intervals of $5$ days between imaging sessions, $4$ sets of images were acquired. Two weeks of aging in mice are approximately equal to one year of aging in humans \cite{dutta2016men}. Datasets were previously morphometrically analyzed using formation and resorption dynamics analysis software \cite{Birkhold2014,Birkhold2014b,birkhold2015monitoring}. Furthermore, previous analysis showed that repeated radiation (4 scans) did not effect bone microstructure \cite{Willie2013}. The scanner was calibrated weekly against a hydroxyapative (HA) mineral phantom; and monthly for determining in-plane spatial resolution. Animal experiments were carried out according to the policies and procedures approved by the local legal representative (LAGeSo Berlin, G0333/09). 

\subsection{Image Preprocessing Pipeline}
Raw data were reconstructed using standard filtered backprojection implemented in the software of the scanner. Resulting images were cropped to contain the tibia and fibula in independent image sets (Fig. \ref{fig:1}C). These images  varied considerably  in size, location and orientation of the bone. Therefore,  a  preprocessing  pipeline that standardizes  the images  is  essential  for training a deep learning  model.  
The first step of this pipeline normalizes the sizes  of the  input  images, in which, the algorithm determines the maximum extension in lateral-medial and anterior-posterior direction as well as the most distal bone part inside the image. Second,  preserving  their  aspect  ratios, a padding of the images in lateral-medial and anterior-posterior direction is performed (Fig. \ref{fig:1}D). Therefore,  a stripe of additional voxels with the same gray values is placed at the border of the image on the padded plane perpendicular to the padding direction. Next, images are cropped to the minimum $z$-stack number (of all the images) from distal direction. 
The $3D$ images are projected in medial-lateral direction onto the anterior-posterior / distal-proximal plane ensuing a $2D$ image,  which, due to the medial-lateral symmetry, nullifies the symmetry-related skeletal difference between the left and right tibia and fibula. After pre-processing, all 2D images are $733$ ($y$) by $161$ ($z$) pixels in size (Fig. \ref{fig:1}E). 

\subsection{Assigning Datasets}
Datasets were separated into three groups: 1) A training and validation set, 2) a test set to further eliminate the possibility of overfitting  of the trained model, and  3) an application set. The training and validation sets consists of $71$ images from $18$ mice at all time points of the right control tibia. The test set contains $8$ images of two randomly selected mice separated from the training and validation set ($2$ image per time point). The application set contains $78$ images of the left loaded tibia and fibula. 

\subsection{Data Augmentation}
DNNs require a large amount of training data  for  stable  convergence  and  high  classification  accuracy. We therefore performed data augmentation, where  we synthetically  increase the size of the training set with geometric  transformations. As suggested by \cite{krizhevsky2012imagenet, simard2003best}, images of the training and validation sets are augmented by applying rotations ($-15\degree$ to $+15\degree$; $1\degree$ steps) and translations ($-22\%$ to $+22\%$; $2\%$  steps in both directions) to increase training accuracy and further prevent overfitting. Maximum augmentation values were chosen to cover potential occurring deviations between images due to the imaging setup. This results in a total  of $23430$ images. Last, images are randomly separated into training and validation sets containing $80\%$ and $20\%$ of the augmented images, respectively. 

\subsection{Bone Age Assessment Model (BAAM) Network}
\label{subsec-training}
A customized DNN consisting of $7$ layers with four convolutional, two pooling and one fully connected layer (Fig. \ref{fig:1}F) was  designed. The output layer consists of four classes: day $0$, day $5$, day $10$, day $15$. Its performance compared to other possible architectures was evaluated (Supplemental data \ref{appendix}). 

By passing an image through the convolutional layers, feature-maps of the image are produced at which the position of the encrypted patterns in kernels are accentuated. This leads to extraction of hidden features of the structure. The deeper the image goes through the network, the more complicated patterns are recognized by the network to perform a classification. At each convolutional layer, the feature map extraction is performed by activating the neurons with the rectified linear function and adding a set of bias terms. These weights and bias values are optimized at each training iteration to provide a higher accuracy. 

Feature-maps are down-sampled after the $2^{nd}$ and $4^{th}$ convolutional layer with a window size and stride equal to $2$. The kernels in all convolutional layers are $3*3$ with a stride of $1$. In the consecutive convolutional layers, $4, 8, 16$ and $32$ feature maps are computed, respectively. We flatten the activation of the last convolutional layer into a vector and pass it to the ending fully connected layer with $32$ and $4$ features from which the last one represents the $4$ age classes. At last, a softmax function is applied on the flattened layer to calculate the probability distribution for each age class.

\subsection{Training Algorithm}
We trained the model with the Adam optimization algorithm \cite{kingma2014adam}. The network is initialized with a truncated normal distribution function (standard deviation: $0.1$). Training is carried out for $7000$ iterations with a batch size of $100$ images. At every $500^{\rm th}$ step the model is applied on a batch of $200$ images of the validation set. The initial learning rate is set to $0.1$, then an exponential decay at every $25^{th}$ step with a $0.96$ rate is performed. Implementation, training, validation and testing of the network was performed using Google TensorFlow \cite{abadi2016tensorflow} on a computer with a single Nvidia Geforce GTX 1070 GPU.

\subsection{Network Performance Evaluation and Validation}
Architecture and hyper parameters of  BAAM network are chosen based on its accuracy in age prediction in validation and test sets. Based on a sensitivity analysis (\ref{appendix}), the best performing DNN was chosen and its age prediction performance was validated by determining the accuracy of the network at three groups to  i) verify the capability of BAAM to perform an end-to-end age prediction based on 3D $\mu$CT image data, ii) eradicate the possibility of overfitting during prediction, iii) demonstrate the capability of transferring the age prediction capability from right to left tibia and fibula, and iv) demonstrate the robustness of the model to predict the age of mice that it has not been trained with:
1) $100$ randomly selected images of the validation set,
2) the $8$ images of the test set, which contains only images of mice it has not seen before,
3) all (n=$20$) images of day 0 of the loaded left tibia and fibula of the application set. These samples represent physiologically loaded tibia and fibula, as at day $0$ of the experiment the left limb has not been subjected to loading treatment yet. Confusion matrices (CM) are determined for all three evaluations, further sensitivity was calculated for all time points.

\subsection{Analysis of Key Skeletal Maturation Regions and Features}
The trained BAAM network (after last training and validation step) applied to the only physiological loaded right tibia (each time point) is further evaluated to identify key regions and features describing the age of the bones. To determine the regions in the images that the network is focusing on for age prediction, saliency maps are calculated. Respectively, a backpropagation is calculated for computing the vanilla gradients \cite{simonyan2013deep,erhan2009visualizing}. The loss gradient is additionally backpropagated to the input data layer. By taking the L1-norm of the loss gradient of the input layer, the resulting heat map intuitively represents the importance of each pixel for age prediction. These maps convey the locations in the image at which the network focuses to predict the age of the bone. At last, saliency maps are normalized to a $[0-1]$ range to enable comparability between different images. 

\begin{figure}[htbp]
\centering
\includegraphics[width=90mm]{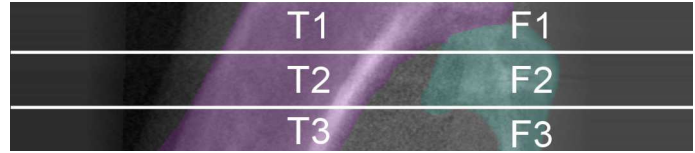}
\caption{Extracted labels for tibia (purple) and fibula (green) and the 6 regions with equal proximal-distal height on each label. T1-T3 and F1-F3 regions belong to tibia and fibula, accordingly.}
\label{fig:3}
\end{figure}

To determine the spatial localization of attention of the network in the process of age assessment, six subregions were defined within the proximal tibia and fibula. Therefore, first the tibia and fibula were manually segmented. Next, these two labels were further divided into $3$ regions with the same heights ($0.56$ mm), o.e., proximal (T1, F1), middle (T2, F2), and distal (T3,F3), cf., Fig. \ref{fig:3}. The summation of intensity values of the saliency map in each region normalized to the summation of intensity  values of the saliency map in the bone region (tibia and fibula) are defined as a measure to indicate the importance of each region for the age estimation (Attention, $[0-1]$).

\subsection{Predicting Rejuvenation Effects on Skeletal Age}
\begin{figure}[htbp]
\centering
\includegraphics[width=90mm]{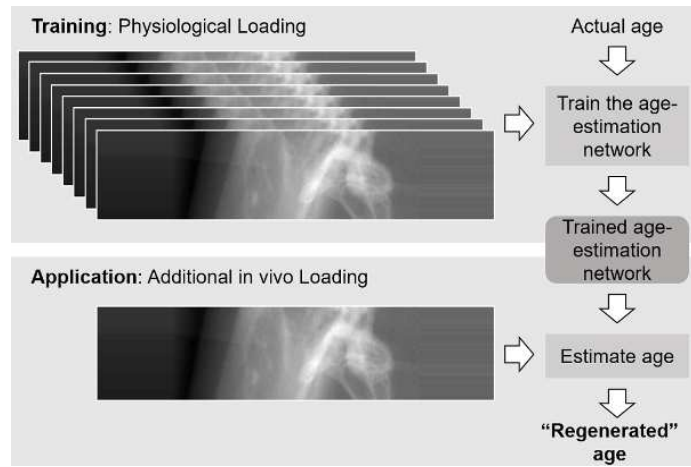}
\caption{First, the network is trained on physiologically loaded bones to predict age (top). Second, the trained network is applied on images of bones treated with additional loading to investigate the rejuvenation effects of treatment (bottom).}
\label{fig:2}
\end{figure}

The bone age prediction model is further applied to the application set (Fig. \ref{fig:2}). The predicted age of each bone at each time point is compared to the actual bone age to investigate rejuvenating effects of load-induced bone (re)modeling on skeletal age. Rejuvenation is defined as the delta age predicted at day 0 and day x divided by the delta time between day 0 and day x. Key regions and features describing the estimated rejuvenated age of the bones are determined (saliency maps). 

\subsection{Correlations to 3D Bone Volume Changes}
A set of 104 3D images were Gaussian filtered and binarized using a global threshold of 273/1000 (456 mg HA/cc), which was determined based on the grey value histogram of the whole ROI.
Fibula and tibia were manually separated, automated segmentation was performed to separate trabecular  and cortical bone regions of tibiae, as described earlier \cite{birkhold2015monitoring}. 
Total tibia bone volume (tibia BV, $\mu m^3$), tibia trabecular bone volume (tibia tb.BV, $\mu m^3$), tibia cortical bone volume (tibia ct.BV, $\mu m^3$), and fibula bone volume (fibula BV, $\mu m^3$) are determined. Correlations between bone volumes and real/predicted age are determined. 

\subsection{Statistical Analysis}
The effect of loading (left loaded tibia, right control tibia), region (tibia: T1, T2, T3; fibula: F1, F2, F3) and time point (day $0$, $5$, $10$, $15$) as well as interactions between terms was assessed using repeated measures ANOVAs. Differences between actual bone age and predicted bone age as well as between loaded and control bones were assessed by paired Student’s t-tests. Values are presented as mean$\pm$standard deviation and statistical significance was set at $p<0.05$.

\section{Results}
\subsection{Performance Evaluation and Bone Age Prediction}
\begin{figure}[htbp]
\centering
\includegraphics[width=\textwidth]{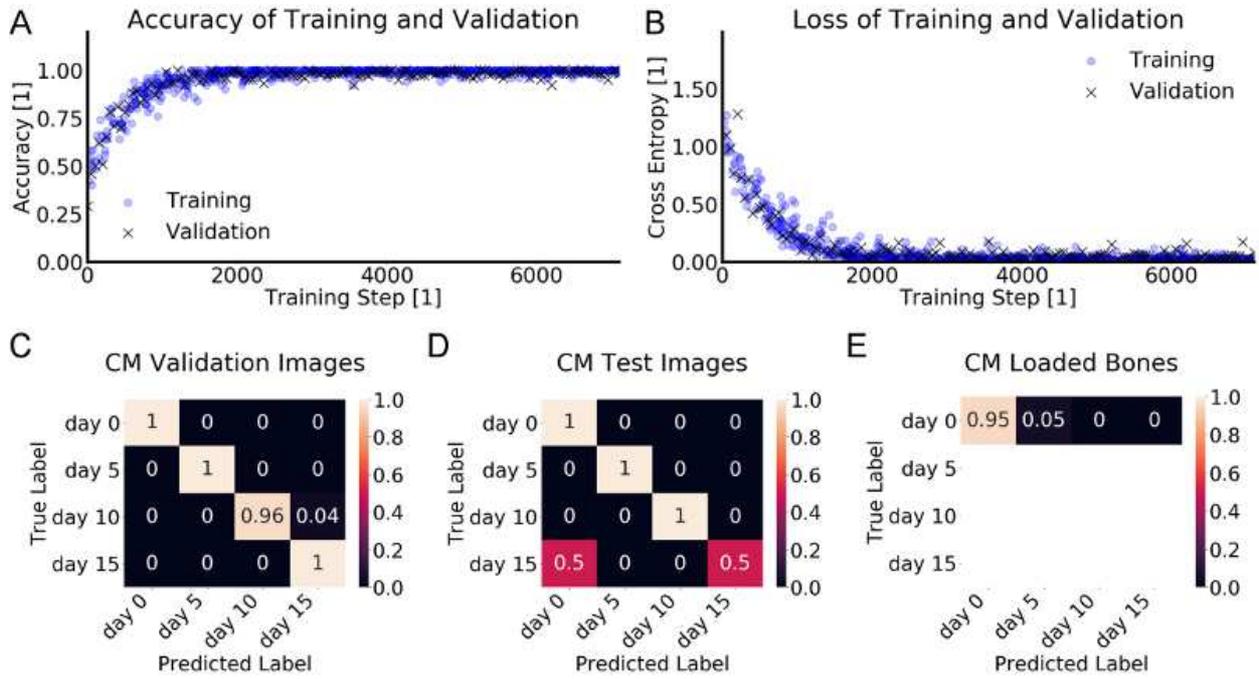}
\caption{Age prediction results and validation. (A) Accuracy of age prediction on training set (blue $\circ$) and validation set (black $\times$) vs. training step. (B) Calculated loss value of age prediction for training set (blue $\circ$) and validation set (black $\times$) vs. training step. Accuracy and loss of age prediction on the validation set is calculated at every $50^{th}$ training step. (C) Confusion matrix of applying BAAM on validation set. (D) Confusion matrix of applying BAAM on test set. (E) Confusion matrix of applying BAAM on images with age of day $0$ of application set. (C-E): Predicted age of samples are compared to their actual age. Values are normalized by the number of images per age class.}
\label{fig:4}
\end{figure}

After $1000$ training iterations, the accuracy of age prediction (training and validation sets) reached $92\%$ and $93\%$ with a loss of $0.21$ and $0.20$, respectively. After $7000$ iterations, the accuracy and loss values for training and validation sets were $100\%$, $99\%$, $0.02$ and $0.03$, respectively (Fig. \ref{fig:4}A-B). Therefore, the network was considered as trained after $7000$ iterations.

For $100$ randomly selected images of the validation set ($21$, $29$, $26$ and $24$ images of days $0$, $5$, $10$ and $15$, respectively), the network correctly predicted all classes except for day $10$, where $96\%$ were assigned correctly to day 10 and $4\%$ were assigned to day $15$. The resulting confusion matrix for comparison of predicted and true age of the images is almost completely diagonal (Fig. \ref{fig:4}C). In the test set, $7$ out of $8$ images were correctly predicted (Fig. \ref{fig:4}D). Only one image of day 15 was wrongly classified as day 0. In the group of images of the left tibia acquired at day $0$, $19$ out of $20$ images were predicted correctly, only one image was wrongly classified as day $5$, resulting in $95\%$ accuracy (Fig. \ref{fig:4}E). With these accuracy values above $95\%$ in all evaluations, we consider our network performing sufficient for age-prediction. 

Accuracy in age prediction for validation and test sets of five similar, evaluated network architectures compared to BAAM are detailed  in the supplemental data (\ref{appendix}).

\subsection{Decoding Bone Tissue Maturation Process}
\begin{figure}[htbp]
\centering
\includegraphics[width=\textwidth]{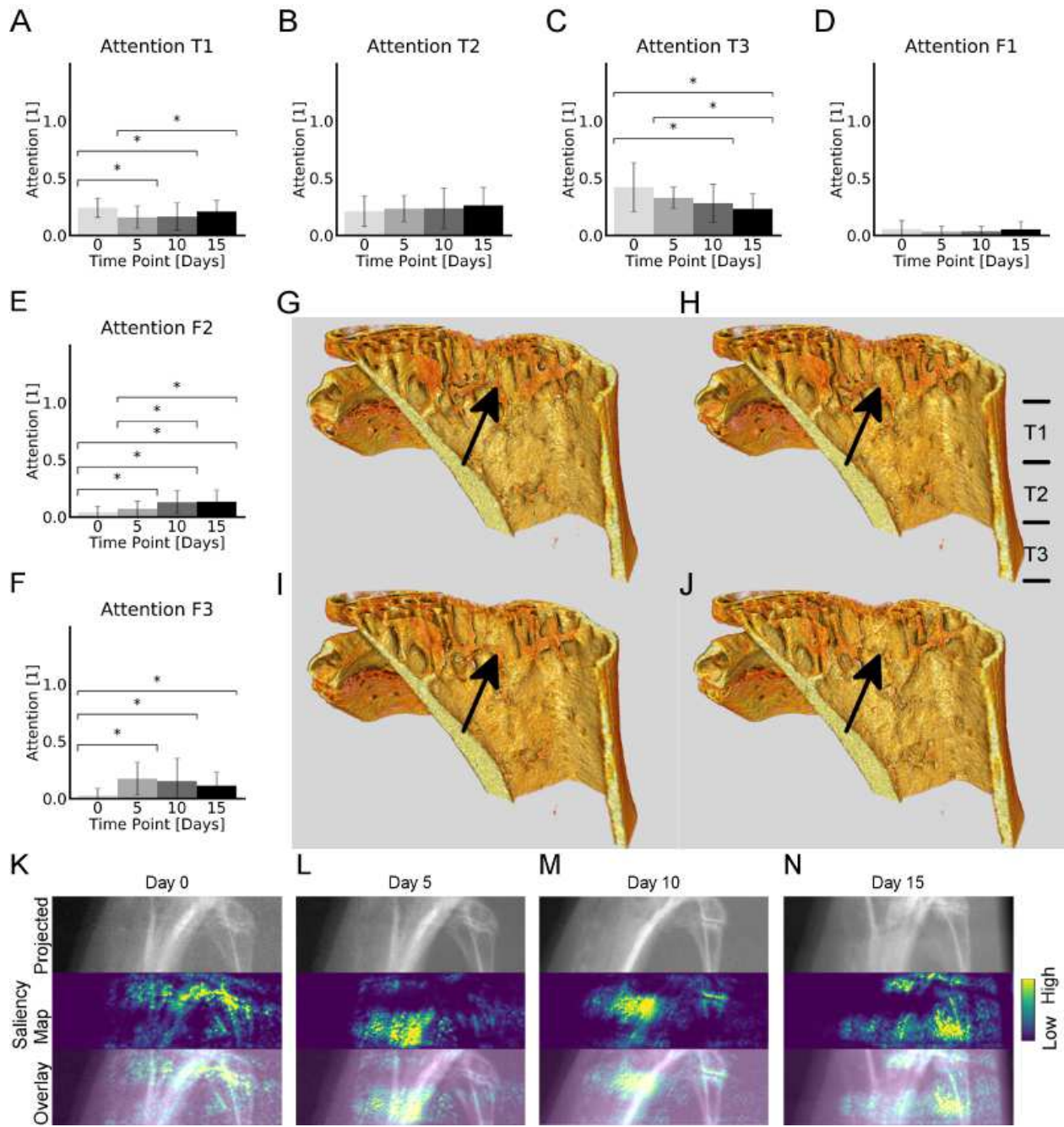}
\caption{Maturation process.(A-F) Temporal changes in attention devoted by network to the six regions for age classification in control group. (A-C) Proximal tibia T1, T2, and T3, (D-F) Proximal fibula F1, F2, F3. Data shown as mean $\pm$ standard deviation. * indicates a significant difference. (G-J) 3D visualization of the same tibia and fibula with a longitudinal cut exposing bone micro-architecture for day 0, day 5, day 10 and day 15. The arrow indicates one area in which  microstructural remodeling occurred over the 15 days. (K-N) Samples of projected images (gray scale image, top), saliency map (color coded from blue to yellow indicating low to high attention, middle) and the overlay of projected image and its saliency map (bottom). (K) day $0$, (L) day $5$, (M) day $10$, (N) day $15$.}
\label{fig:5}
\end{figure}

Saliency maps were calculated for all correctly classified  images. 
%comparing regions
One further image was excluded due to a different orientation of the bone.
The different regions received different amounts of attention from the network during the process of age estimation (ANOVA; $p<0.01$; Fig. \ref{fig:5} A-F). Comparing the network attention devoted to different regions of the bone revealed, that at day 0 ($42\%\pm22\%$) and day 5 ($33\%\pm9\%$) T3 received significantly higher attention than all other regions ($p\leq0.01$). At day 10, T2 and T3 ($p\leq0.04$) and at day 15 T1, T2 and T3 ($p\leq0.02$) received significantly higher attention than the other regions. In general, all tibial regions received at day 0 and day 15 higher attention than all fibular regions ($p\leq0.03$). At day 5 and day 10, only T3 received a higher attention than all fibular regions ($p\leq0.04$). 

%changes over time
The attention of the network to the different regions changed with time (ANOVA; $p<0.01$). In the tibia, the attention on T1 for age prediction is on day 0 ($24\%\pm9\%$) significantly higher than on day 5 ($16\%\pm10\%$; $p<0.01$) and day 10 ($16\%\pm12\%$, $p=0.01$). At day 15, only a trend could be identified ($21\%\pm9\%$; $p=0.14$; Fig. 5A). 
T2 receives similar attention at all time points; $21\%\pm14\%$, $23\%\pm11\%$, $24\%\pm18\%$, and $26\%\pm16\%$.  In regions T3 a continuous decrease in attention during maturation takes place ($42\%\pm22\%\rightarrow 23\%\pm13\%$), leading to a significant reduced attention from day 10 onward ($p\leq0.01$). In the fibula, attention to the F1 regions remains small throughout the 15 days ($5\%\pm8\%$, $3\%\pm4\%$, $4\%\pm4\%$, $5\%\pm7\%$). The attention to F2 significantly increases with maturation ($4\%\pm5\%\rightarrow 13\%\pm11\%$, $p\leq0.01$).
It also significantly increases in each time step except the time step from day 10 to 15 (day $0\rightarrow5\rightarrow10\ : 4\%\pm5\%\rightarrow7\%\pm7\%\rightarrow13\%\pm10\%$, $p\leq0.02$). 
In 
region F3, attention jumps from $3\%\pm6\%$ at day 0 to $17\%\pm14\%$ at day 5 ($p<0.01$). 
Afterwards a slight drop from $15\%\pm20\%$ (day 10) to $11\%\pm12\%$ occurs at day 15 
($p=0.08$). 
A cross-sectional cut through a 3D representation of one bone at each  time-point shows  similarities and changes of the structures (Fig. \ref{fig:5}G-J). 
Representative saliency maps are shown in (Fig. \ref{fig:5} K-N). Qualitative analysis  of the attention devoted to tibia and fibula reveals that the network focuses on clusters of pixels for its analysis. 
%Can you say how many pixels constitutes a cluster
% @Bettina: It is difficult to say since it is different from image to image. Here I meant the network focuses on groups of neighboring pixels instead of single pixels.
\subsection{Identifying Rejuvenation Effects of in vivo Loading}
\begin{figure}[p]
\centering
\includegraphics[width=\textwidth]{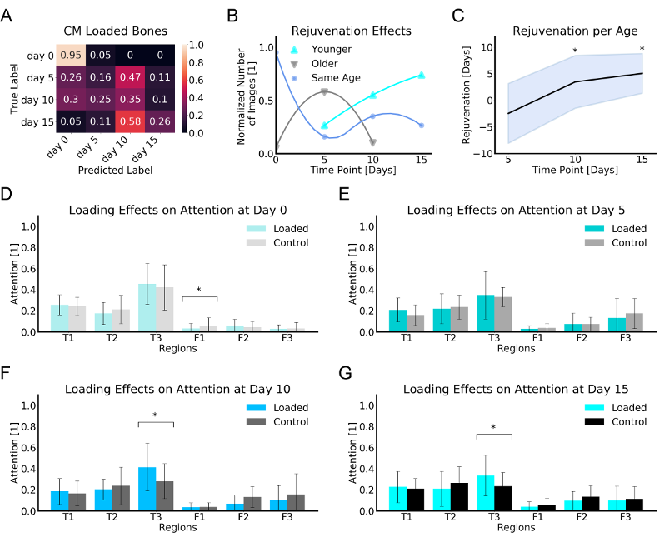}
\caption{Rejuvenation process. (A) Confusion matrix of applying the BAAM network on application set consisting of images of bones treated with extra loading. Predicted age of samples are compared to their actual age. Values are normalized by the number of images per age class. (B) Rejuvenation effect of extra loading treatment vs loading duration. Normalized number of rejuvenated images vs loading period: green. Normalized number of images classified with their actual age vs loading period: gray. Normalized number of images classified older than their actual age vs loading period: blue. The lines are $3^{rd}$ degree polynomial fits to the scattered data. (C) Mean rejuvenation per age group with standard deviation as highlighted area. (D-G) Attention in proximal tibia (T1-T3) and fibula (F1-F3) in control and loaded bones. (D) Day 0. (E) Day 5. (F) Day 10. (G) Day 15. Values are shown as mean$\pm$standard deviation. * indicates a significant differences between loaded and control bones (student's t-test, $p<0.05$).}
\label{fig:6}
\end{figure}

To study potential rejuvenation effects of mechanical loading on bones, $78$ images of bones subjected to in vivo loading are analyzed (application set). At day $0$, $95\%$ of bones were classified with their actual age. For the loaded bones at day 5, 10, and 15, predicted age differed noticeably from actual age (Fig. \ref{fig:6}A). After $5$ days of loading, $47\%$ of the images were classified as being 5 days older, $26\%$ as being 5 days younger, and only $16\%$ were identified with their actual age.
After $10$ days of loading, $55\%$ of the bones were classified as younger than their actual age ($30\%$ as day $0$ and $25\%$ as day $5$). A total of $35\%$ were identified with their actual age and  $10\%$ were classified as to be older. After $15$ days of loading, $74\%$ of the bones were classified as younger than their actual age. For $5\%$, BAMM predicted that the images belong to mice that were 15 days younger,  $11\%$ appeared to be 10 days younger and $58\%$ were classified as $5$ days younger than their actual age. Only $26\%$ of the bones were classified by its actual age, i.e., day $15$.

Investigating the rejuvenation effects during the course of loading (Fig.~\ref{fig:6}B, green line) reveals that with increasing duration of loading (day $5$ $\rightarrow$ day $10$ $\rightarrow$ day $15$) higher percentage of images are classified younger than their actual age ($25\%\rightarrow55\%\rightarrow74\%$). On the other hand, $5\%$, $58\%$, and $10\%$ of samples at day $0$, $5$ and $10$, respectively were classified older than their actual age. % It is interesting that 5 days overestimates bone age compared to  0.
The amount of samples categorized with their actual age increases from day $5$ to day $10$ but decreases as loading treatment continues ($16\%$, $35\%$ and $26\%$ for  day $5$, $10$, and $15$, respectively). Resulting mean assigned age was $7.63\pm5.62$ at day 5,   $6.58\pm5.01$ at day 10 and  $10.26\pm3.90$ at day 15. This results in rejuvenation effects of $-2.37\pm5.62$ at day 5,  $3.68\pm4.96$ at day 10 ($p<0.01$) and $5.00 \pm 3.73$ at day 15 ($p<0.01$; Fig. \ref{fig:6}C).

Loading affected the attention of the DNN network (ANOVA; $p<0.01$). Additionally, the distribution of attention between regions was affected by loading (ANOVA; $p<0.01$). Comparison of attention in different regions between loaded and control limb reveals that -- with the exception of F1 ($p=0.02$) --  there is no significant difference in attention in different regions between the left and right limbs at day 0  (Fig.~\ref{fig:6}D). After 5 days of loading; there is no significant difference in attention in T1-T3 and F1-F3  (Fig. \ref{fig:6}E). After 10 and 15 days of loading; however, the attention in T3 regions is significantly higher in loaded compared to control limbs  ($28\%\pm17\%$ vs. $41\%\pm23\%$, $p=0.01$ and $23\%\pm13\%$ vs. $33\%\pm19\%$, $p=0.04$ at days 10 and 15, accordingly,  (Fig. \ref{fig:6}F-G).

%  This difference might be related to presence of paw preference (called handedness) in mice \cite{biddle1993genetic}.
\subsection{Decoding Rejuvenation Effects of in vivo Loading}
\begin{figure}[p]
\centering
\includegraphics[width=\textwidth]{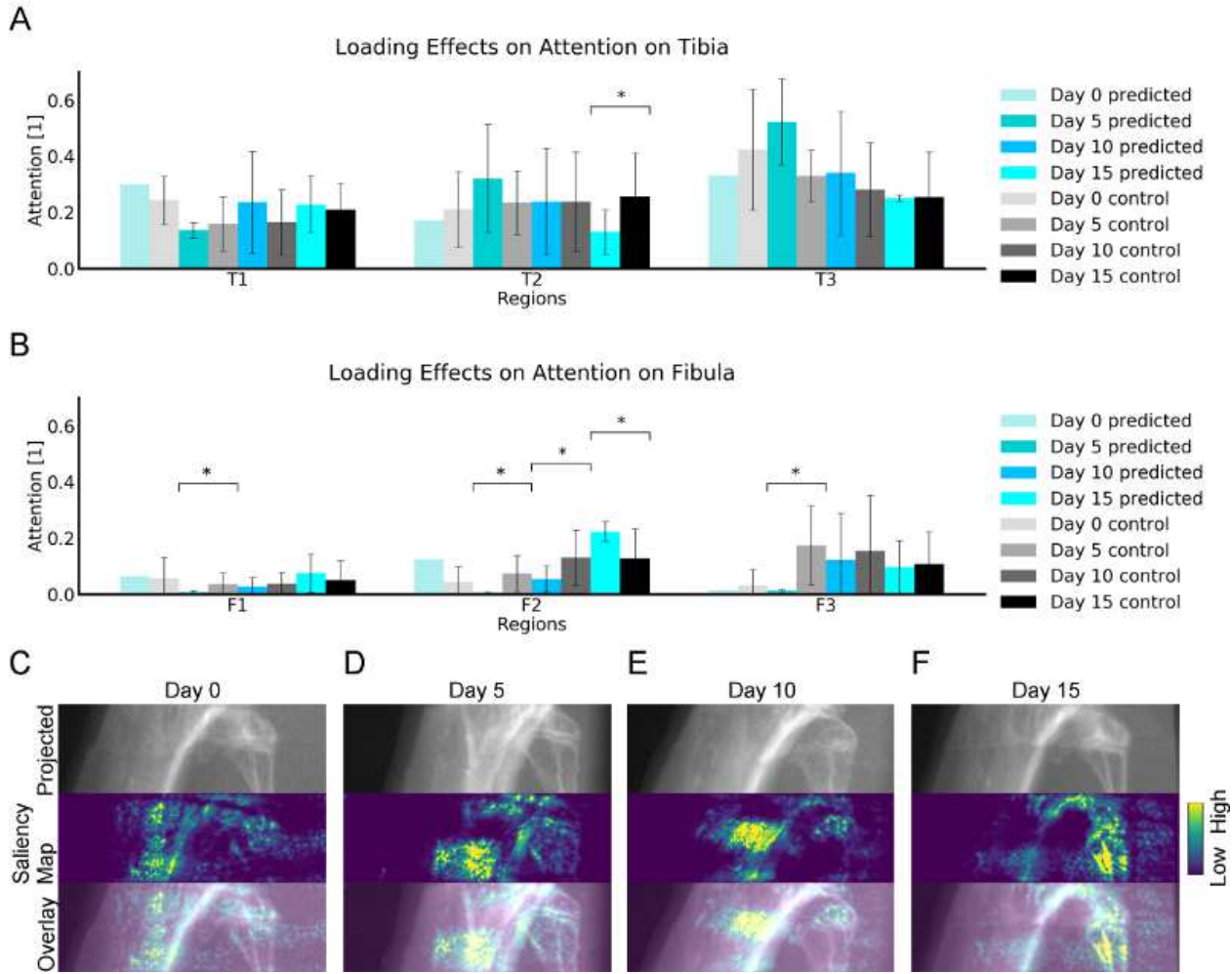}
\caption{Rejuvenation process. (A, B) Attention at bones subjected to 15 days of loading and classified as day 0, 5, 10, or 15 compared to attention of these groups in the control limbs. (A) Attention to T1-T3 regions (B) Attention to F1-F3 regions. Data shown as mean $\pm$ standard deviation. * indicates a significant difference (student's t-test, $p<0.05$). (C-F) Samples of projected images (gray scale image, top), saliency map (color coded from blue to yellow indicating low to high attention, middle) and the overlap of projected image and its saliency map (bottom) from loaded group. (C) day $0$. (D) day $5$. (E) day $10$. (F) day $15$. }
\label{fig:7}
\end{figure}

To analyze the importance of each region within the process of rejuvenation, saliency maps of images of day 15  of loaded bones predicted as days 0, 5, 10, and 15 were compared to the distribution of attention in control image of each of these specific time points (Fig. \ref{fig:7}A-B). 
After 15 days of loading, the region T1, T2, and T3 received $23\%\pm14\%$,  $21\%\pm16\%$, and T3 $33\%\pm19\%$ of the attention, respectively. F1, F2, adn F3 received $4\%\pm5\%$, $10\%\pm9\%$, and $10\%\pm13\%$ of the attention, respectively.
In all tibial regions, no significant difference between bones predicted younger and control bones of these specific ages were found for any of the four time points. The single bone (imaged at day 15) predicted as day 0 received $30\%$ attention at T1 (control: $24\%\pm9\%$),  $17\%$ at T2 (control: $21\%\pm14\%$) and $33\%$ at T3 (control: $42\%\pm22\%$). 
The bones predicted as day 5 received $14\%\pm3\%$ at T1 (control: $16\%\pm1\%$; $p=0.24$), $32\%\pm19\%$ at T2 (control: $23\%\pm12\%$; $p=0.32$) and $52\%\pm15\%$ at T3 (control: $33\%\pm10\%$; $p=0.16$). 
The bones predicted as day 10 received $23\%\pm18\%$ at T1 (control: $16\%\pm12\%$; $p=0.13$), $23\%\pm19\%$ at T2 (control: $24\%\pm18\%$; $p=0.45$) and $34\%\pm22\%$ at T3 (control: $28\%\pm17\%$; $p=0.23$). 
The bones predicted as day 15 received $23\%\pm10\%$ at T1 (control: $21\%\pm10\%$; $p=0.38$) and $25\%\pm10\%$ at T3 (control: $25\%\pm16\%$; $p=0.22$).
Only for the $5$ bones, which were predicted as a loaded bone of age day 15, i.e., no effect of  the loading was observed, a significant difference with respect to the control bones being imaged at day 15 could be identified at the T2 region ($13\%\pm8\%$, vs. control: $25\%\pm16\%$; $p=0.01$). 

The single bone predicted as day 0 received $6\%$ attention at F1 (control: $6\%\pm9\%$),  $12\%$ at F2 (control:  $4\%\pm5\%$) and $1\%$ at F3 (control:  $3\%\pm6\%$). 
The bones predicted as day 5 received significant less attention at all fibular regions: F1 $1\%\pm1\%$ (control: $3\%\pm4\%$; $p<0.01$), F2 $0\%\pm0\%$ (control: $7\%\pm7\%$ ; $p<0.01$) and F3 $1\%\pm0\%$ (control: $17\%\pm15\%$; $p<0.01$). 
The bones predicted as day 10 received at F1 and F3 comparable attention as the control ones of day 10  ($3\%\pm3\%$ control: $3\%\pm4\%$; $p=0.22$; $12\%\pm15\%$; control: $15\%\pm20\%$; $p=0.32$). F2 received less attention than the control bones of this age ($5\%\pm5\%$; control: $13\%\pm10\%$; $p<0.01$). 
The bones predicted as day 15 received at F1 and F3 attention that is comparable to the control ones of day 15 ($7\%\pm7\%$ control: $5\%\pm7\%$; $p=0.24$; $10\%\pm10\%$; control: $11\%\pm12\%$; $0.42$) and
F2 received more attention than the control bones of this age ($22\%\pm3\%$; control: $13\%\pm10\%$; $p<0.01$; Fig. \ref{fig:7}B). Saliency  maps  for selected  images from  each  of  the four time points are given in Fig. \ref{fig:7}C-F.

In summary, distribution of  attention was in  $89\%$ of the after 15 days analyzed loaded left tibia comparable to the right control of the time point (day 5, 10 or 15) they were predicted to resemble. In the fibula, $56\%$ of comparison cases showed no significant difference (9 comparison cases per tibia/fibula = 3 regions $\times$ 3 time points). This further confirms that the load-induced bone (re)modeling in these mice generates similarities between older loaded bones and younger control bones, therefore a rejuvenation effect of loading can be concluded. 

\subsection{Correlations between Rejuvenation and Morphological Changes}
The tibia and fibula total bone volume and cortical bone volume did not change significantly within the 15 days (Table \ref{table:1}). Tibial trabecular bone volume decreases significantly from day 0 ($0.30\pm0.07$) to day 15 ($0.24\pm0.05$; $p<0.01$). Loading significantly affected total tibia bone volume (all time points; $p\leq0.01$), cortical tibia bone volume (all time points; $p\leq0.02$) and trabecular tibia bone volume (day 0 and 15; $p\leq0.01$, Table \ref{table:2}). A significant, but weak correlation was found between Tibia BV in loaded limbs and real age, Tibia ct.BV and real age, and Tibia ct.BV and estimated age (Table \ref{table:3}). Additionally, trabecular bone volume (Tibia tb.BV) in control limb correlated significantly, but weak with real age. Trabecular bone volume (Tibia tb.BV) in loaded limb correlated weak, but significantly with estimated age (Table \ref{table:3}). 

\begin{table}[ht]
\small 
\centering
\begin{tabular}{ |p{4cm}||c|c|c|c|}
 \hline
 \multicolumn{5}{|c|}{Morphometry Control Bones} \\
 \hline
 & day 0 & day 5 & day 10 & day 15 \\
 \hline
Fibula BV [$mm^3$]  &  $0.78\pm0.16$  & $0.79\pm0.17$&  $0.82\pm0.19$& $0.78\pm0.15$\\
Tibia BV  [$mm^3$]  &  $2.53\pm0.13$  & $2.53\pm0.16$&  $2.54\pm0.14$& $2.50\pm0.12$\\
Tibia tb.BV [$mm^3$] &  $0.30\pm0.07$  & $0.28\pm0.07$&  $0.25\pm0.06$& $0.24\pm0.05$\\
Tibia ct.BV [$mm^3$] &  $2.23\pm0.13$  & $2.25\pm0.13$&  $2.28\pm0.13$& $2.26\pm0.10$\\
 \hline
\end{tabular}
\caption{Morphometry  parameters of tibia and fibula determined based on in vivo $\mu$CT images at day 0, 5, 10, 15  of right control fibula and tibia subjected only to physiological loading. Data shown as mean $\pm$ standard deviation.}
\label{table:1}
\end{table}

\begin{table}[ht]
\small 
\centering
\begin{tabular}{ |p{4cm}||c|c|c|c| }
 \hline
 \multicolumn{5}{|c|}{Morphometry Loaded Bones} \\
 \hline
 & day 0 & day 5 & day 10 & day 15 \\
 \hline
Fibula BV [$mm^3$]  &  $0.83\pm0.15$  & $0.82\pm0.15$&  $0.82\pm0.17$& $0.81\pm0.12$\\
Tibia BV  [$mm^3$]  &  $2.69\pm0.18^\ast$  & $2.74\pm0.17^\ast$&  $2.84\pm0.14^\ast$& $2.86\pm0.14^\ast$\\
Tibia tb.BV [$mm^3$] &  $0.33\pm0.06$  & $0.31\pm0.06$&  $0.31\pm0.06^\ast$& $0.33\pm0.06^\ast$\\
Tibia ct.BV [$mm^3$] &  $2.36\pm0.19^\ast$  & $2.43\pm0.19^\ast$&  $2.52\pm0.14^\ast$& $2.53\pm0.15^\ast$\\
 \hline
\end{tabular}
\caption{Morphometry parameters of tibia and fibula determined based on in vivo $\mu$CT images at day 0, 5, 10, 15  of left fibula and tibia subjected to additional axial compression loading. Data shown as mean $\pm$ standard deviation. $^\ast$ indicates a significant difference between loaded and control bones (student's t-test, $p<0.05$).}
\label{table:2}
\end{table}

\begin{table}[h!]
\small 
\centering
\begin{tabular}{ |p{4cm}||c|c|c|c|c|c|}
 \hline
 \multicolumn{7}{|c|}{Correlations Bone Volume vs. Time} \\
 \hline
 & \multicolumn{4}{|c|}{Loaded} & \multicolumn{2}{|c|}{Control}  \\
\hline 
 & \multicolumn{2}{|c|}{ Real age} & \multicolumn{2}{|c|}{Predicted age} & \multicolumn{2}{|c|}{Real age} \\
\hline 
 & $R^2$& F &$R^2$& F& $R^2$& F\\ \hline 
Fibula BV [$mm^3$]  &0.01      &0.59          &0.02  &0.29          &0.00  &0.96 \\
Tibia BV  [$mm^3$]   &\textbf{0.15}   &0.01$^\ast$     &0.05  &0.11          &0.00  &0.65 \\
Tibia tb.BV [$mm^3$] & 0.00     & 0.98          &\textbf{0.10}  & $0.03^\ast$    &\textbf{0.12}  &$0.02^\ast$\\
Tibia ct.BV [$mm^3$] & \textbf{0.14}   &$0.01^\ast$     &\textbf{0.11}  &0.02$^\ast$     &0.01  &0.43\\
\hline
\end{tabular}
\caption{Regression analysis: Correlations between morphology (bone volume) and time points (real age and predicted age). Weak correlation are shown in bold, $^\ast$ indicates significance of correlations ($p<0.05$).}
\label{table:3}
\end{table}

\section{Discussion}
\label{Discussion}

It is known that bone fracture resistance increases with skeletal maturation and aging, as well as in response to certain therapy. Whole bone fracture resistance is determined by bone quantity, which encompasses geometric, microarchitectural, and material properties (i.e., trabecular architecture, mineralization,  crosslinking,  microcracks). Little is known about the interplay between all of these properties/factors contributing to compromised or recovered bone quality. Most previous studies have focused on individual  features, while lacking global optimization, as individual contributions of static  (e.g. trabecular bone volume or bone mineral density distribution) or dynamic (e.g. bone formation rate) features. 
Here, we propose a deep learning approach to tackle this challenge by creating a model that utilizes the complete content of X-ray attenuation images. We  developed a $\mu$CT image-based  method enabling an end-to-end age prediction with high accuracy, that allows identifying bone sub-regions relevant for this classification. We utilized the developed method to study skeletal tissue maturation and the localized rejuvenation effects of in vivo dynamic controlled loading on mouse bone tissue age. 

Our results show that complex processes, such as bone tissue maturation and adaptation, can be reconstructed by in vivo imaging-based deep learning and quantitative analysis of learned features. Even subtle changes, as occurring during one remodeling cycle of mice bones \cite{birkhold2015monitoring}, could be identified, as our model predicts four time points between 26 and 28 week old mice based on in vivo $\mu$CT images with an accuracy of more than $95\%$. This classification is presumably linked to bone mass, shape, micro-architecture and material properties alteration through different length scales, as all these processes change in a site-specific manner during growth, maturation and aging. These changes take place by modeling, remodeling, mineralization and demineralization processes \cite{parfitt2001skeletal,Birkhold2014,buenzli2018late,lukas2013mineralization,somerville2004growth,lynch2011tibial} and are directly or indirectly encoded in the image created by interaction of photons with matter, as the resulting attenuation is, besides bone mass in the photon beam, dependent on the local atomic number, and therefore calcium content. 

Additionally, we show a link between the age prediction dynamics to age-related trabecular bone loss occurring between the age of 26 and 28 weeks \cite{Willie2013}, which is in accordance with other studies reporting a loss in trabecular bone volume in C57Bl/6 mice from 26 weeks onward \cite{glatt2007age}. Therefore, these mice are expected to be in a phase of starting trabecular bone loss.  However, in the proximal fibula, we observed an increase of cortical bone volume in the same mice between 19 to 22 weeks \cite{moustafa2009mouse}. Unfortunately, this and other studies investigated the fibula bone volume changes only in young mice during skeletal maturation \cite{moustafa2009mouse,ko2011deterioration}. In adult humans, higher deterioration of the tibia than the fibula with aging has been reported \cite{mcneil2009geometry}. Here we detected no significant changes in the fibula bone volume, therefore also no correlations between age prediction and fibula bone volume. However, it has to be taken into consideration, that our volume of interest was centered on the proximal tibia.

Here, we further showed, that localization of aging-related information in the bone identified by the model can be extracted in a quantitative manner. This  localization of age-information effects, here quantified by attention distribution of the network extracted using a saliency map post-processing,  differed between regions and was affected by time. The age information seems to be more manifested in tibia than in fibula, as at all time points more than $68\%$ of the attention of the network is focused on the tibia. In line with this, greatest structural changes could be identified in the trabecular region of the tibia, with a $14\%$ loss of bone volume. This trabecular loss has been previously quantified in more detail \cite{Birkhold2014}. In general, the most distal part of the proximal tibia (T3) received the greatest attention, but over time there was decreasing in attention. This is in line with previous described structural changes in the tibia, where a strong loss of trabeculae during adulthood \cite{ferguson2003bone,Willie2013} and a nearly complete disappearance of trabeculae after 78 weeks \cite{Birkhold2014} is reported. The second highest attention was received by the region located closest to the growth plate (T1), where longitudinal growth, and therefore modeling and primary (fast) mineralization occur. This growth persists with aging, although it slows down after puberty \cite{halloran2002changes}. Therefore, it can be assumed, that the network focuses on the regions with the most changes in bone volume and density occurring over the monitored 15 days. In the fibula the attention was the lowest in F1. However, it must be considered, that the most proximal part of the fibula is located below the tibia (see Fig. \ref{fig:5}G-J). Attention to F2 and F3 varied with time, which might be linked to fibular (re)modeling. In general, attention was located in clusters (see Fig. \ref{fig:5}K-N) not individual trabeculae, which let us conclude that mineralization (change in grey values) and (re)modeling together affect the age estimation.

Bones receiving additional loading were estimated to be younger. Already after day 5 of loading, predicted age starts to diverge from actual age, which might suggest a restructuring in response due to the new local loading conditions. From day 10 onward, the bones appear to be significantly younger (4 and 5 days younger after 10 and 15 days of loading, respectively). This might reflect an increase in bone strength after restructuring. However, this needs to be investigated in detail by analyzing orientation of individual trabeculae or in-silico modeling of bone strength.
Previous studies showed in adult C57BL/6 mice adaptive adjustments in the shape of formation and resorption sites at trabecular \cite{Birkhold2014,lambers2015bone} and cortical sites \cite{birkhold2017tomography}, mineralization dynamics \cite{lukas2013mineralization} as well as material properties on a macro- and micro-scale \cite{bergstrom2017compressive} in response to loading. These adaptations have been shown to differ between different bone sites, such as endocortical and periosteal surface \cite{birkhold2016periosteal} or at metaphyseal versus diaphyseal sites \cite{birkhold2017tomography,bergstrom2017compressive}, leading to bone shape adaptation, as e.g. second moment of inertia at proximal metaphysis changes with loading in 22 week old mice \cite{carriero2018spatial}. 
All this information is potentially analyzed in a combined manner by the deep neural network. Here, we established a link to bone volume changes and found a significant correlation between predicted age and cortical as well as trabecular bone volume. Future studies will investigate further correlations, e.g., by quantifying structural changes in sites identified as age-determining. 

To investigate, if the observed restructuring leads to a younger appearance of the bones, we compared the distribution of attention on loaded bones after 15 days, classified younger, to the control bones of the classified ages. The similarities of distribution of attention, together with the comparability of correlations in 1) predicted age and trabecular bone volume in loaded bones and 2) real age and trabecular bone volume in control limbs, lets us speculate, that loading induces tibial restructuring towards a younger appearing bone. This rejuvenation is strongly manifested in the dynamic trabecular  structure. This is also the only bone compartment, where we detected an age-related short-term loss of bone volume in the control limb and a constant bone volume with time in the loaded limb. This finding is further supported by previous studies investigating trabecular bone gain in response to loading \cite{lynch2011tibial,Willie2013,sugiyama2010functional} by comparing  loaded proximal tibia of adult mice with internal nonloaded control limbs. 

In the cortical compartment of the loaded tibia, a bone volume gain was found, which made this bone compartment most similar to the control bones of day 0. This finding is supported by Sugiyama et al., who found similar load-induced cortical volume  gain in the proximal tibia of adult mice \cite{sugiyama2010functional}. In a previous study, we showed that  adaptive bone formation in the lateral metaphyseal region is greater than in the medial region while medial adaptive resorption is greater than in the lateral region. A slight positive anterior and posterior remodeling balance was reported. Further, adaptation occurs by increasing periosteal formation \cite{birkhold2017tomography}. %could you add lateral anterior or posterior to clarify region and sentence below is not clear
This suggest a structural adaptation of cortical bone for the sake of increasing its moment of inertia as a response to the bending force.
In the fibula, regardless of the observed rejuvenation, in $4$ out of $9$ comparisons, significant differences in attention distribution between loaded bones classified younger and the control bones of the classified age was observed. Therefore, since the fibula is in general assumed to be mechano-sensitive \cite{moustafa2009mouse}, load-induced changes seem to trigger a fibula restructuring that does not lead to a younger appearance of the bone. However, it must be considered, that the imaging region was optimized for tibial analysis. Future studies might include a greater fibula region into the analysis as fibula adaptation is in general not well studied. 

To conclude, loading seems to change the appearance of bones towards a younger age. In our study, this effect is greatest in tibiae than in fibulae and mainly manifested in the trabecular region, which is the compartment  most affected by 
bone loss in early osteoporosis \cite{osterhoff2016bone}. Since fragility fractures in relatively young individuals are mainly vertebral compression fractures, therefore "trabecular fractures" \cite{svedbom2014epidemiology}, an adaptation towards a younger trabecular bone might be in combination with an macroscopic adapted cortical shell contributing to reducing fracture risk in early osteoporosis.

Current methods assessing bone maturation and aging mainly focus on specific dynamic features  \cite{burghardt2010reproducibility,bouxsein2010guidelines,mader2013quantitative,birkhold2015monitoring,ruegsegger1976quantification,macdonald2011age}. Recently developed machine learning methods, however, consume the complete image content while performing a classification task. Recently, first applications to preclinical and clinical image data of bone showed, that a machine learning classification into healthy and disease state is achievable \cite{Sharma2016b,Singh2017}. Going from there the  next step, we developed a framework allowing for a tracking of dynamic bone changes in physiological aging/maturation conditions. Additionally, by training with physiologically loaded bones and applying the network on in vivo loaded bones, we demonstrate loading causes bone (re)modeling and we further analyzed the dynamics of this bone (re)modeling, without the need for transfer learning \cite{greenspan2016guest,van2015transfer}. 

Compared to  DNN developed for skeletal bone age assessment in pediatrics \cite{spampinato2017deep,torres2017bone,lee2017fully}, we therefore went further towards a prognostic tool. Saliency maps were previously suggested to qualitatively visualize the importance of different bones in maturation of pediatric hands  \cite{Lee2017a}. Here, we show, for the first time,  a quantitative analysis of the distribution of learned features by employing saliency maps \cite{simonyan2013deep}.  This allows to identify the localization of age-relevant information in bone. Future investigations may analyze the dynamics of aging and treatment-induced regeneration in a site-specific manner. 

Our study has also limitations. First, the chosen samples may not represent a larger population. However, we validated our method in a three-fold manner and received accuracy above $95\%$. Additionally, structural parameters derived from $\mu$CT and histomorphometric indices of bone formation, which we have measured previously in these mice, are similar to those reported in similarly aged female $C57Bl/6$ mice \cite{holguin2013adaptation,lynch2011tibial}. Adhering to the principles of the three Rs, specifically reduction of animal number, motivated us to re-examine these datasets, rather than perform new studies on additional mice. Second, skeletal aging in mice differs to that in humans \cite{jilka2013relevance} and thus, translation of these results to human bone behavior requires further investigation. Since age-related bone loss in humans resembles to a certain extend findings in mice \cite{ahlborg2003bone,szulc2006bone}, it can be speculated that a  future application to human bone might resolve similar patterns.

Given the proven performance on reconstructing bone tissue maturation and adaptation processes, we expect our study to pave the way for future studies to investigate a wide variety of biological processes involving continuous morphological changes. This may include developmental and aging stages, the progression of diseases \cite{yang2017examining,pflanz2017sost} or the response to treatments, and dynamic processes that have often been reduced to binary classification problems. Automated computational analysis, as shown here, could reveal morphological changes at much earlier stages than recognized previously. Therefore, the effects of loading duration on overall bone adaptation might be another future application, as recently shorter loading durations have been suggested  \cite{yang2017effects,sun2018evaluation}. Furthermore, as features can be used to classify biological structures based on morphology \cite{asgharzadeh2018feature}, a combination of the proposed  deep learning and saliency maps approach can lead to more detailed insights into the contribution of individual localized biological processes on the overall adaptation "state" of the bone.  

Here, we described a method for estimating bone age as a surrogate for bone quality in mice imaged with $\mu$CT. Our method can be rapidly translated to clinical applications by examining clinical $\mu$CT (eg. HR-pQCT at a voxel size of 61 microns), which is increasingly used in clinical trials to investigate vBMD and microstructural changes in response to pharmacological treatments \cite{cheung2014effects,tsai2015comparative}. 

\section{Conclusions}
% aging
We combined an experimental study with longitudinal imaging and a deep learning framework. This allowed for a bone (tissue) age prediction and an identification of bone sites that primarily contribute to age classification. Thus for the first time, we could directly track short-time aging in bone in a temporal manner.
% regions: T3
By quantitatively analyzing saliency maps of learned features, we could show that the metaphyseal parts closest and most distant to the growth plate are highly contributing to the temporal age information encoded in bone images during tissue maturation.
% adaptation -> younger appearing bones
We could further show that loading triggers dynamic processes leading to a younger appearance of the bone. More specifically we could temporally quantify these rejuvenating effects, as bone receiving 15 days of loading treatment was classified 5 days younger than the contra-lateral internal control.
% regions: tibia younger, fibula different
We demonstrated that our loading regime induces structural correspondence between younger and older tibiae, while in fibulae, despite causing (re)modeling, no rejuvenation effect could be detected. 
One possible biological interpretation of these findings is that loading recovers the age-related bone loss in the tibia - therefore it rejuvenates, whereas the fibula, which is at the age investigated in the present study does not incur bone loss, and therefore is adapting to the loading in a non-physiological manner (in terms of rejuvenation-related strengthening through bone (re)modeling).  These findings and the introduced method provide an ideal framework to further improve our  understanding of skeletal aging in mice as well as in humans. It further demonstrates that machine-learning based characterization can help to better monitor and understand dynamic changes in bones due to aging and disease and may help to optimize treatments for bone disease such as age-related bone loss and osteoporosis. 

\section*{Acknowledgements}
This study was partially supported by the Cluster of Excellence EXC 2075 “Data-Integrated Simulation Science (SimTech)” and by Shriners Hospital for Children-Canada.
\section*{Conflict of Interest Statement}
Annette I. Birkhold (none), Bettina M. Willie (none), Pouyan Asgharzadeh (none), Oliver Röhrle (none).
\section*{Author’s Roles}
Study design: AB and PA. Data collection: AB, BW. Contributing software tools: PA. Data analysis: PA, AB. AB and PA wrote the manuscript with the help of BW and OR. All authors take responsibility for the integrity of the data analysis.

\singlespace
\bibliographystyle{model1-num-names}

\bibliography{references}

\appendix
\section{Sensitivity Analysis of Network Architecture and Hyper Parameters} 
\label{appendix}
\begin{figure}[htbp]
\centering
\includegraphics[width=140mm]{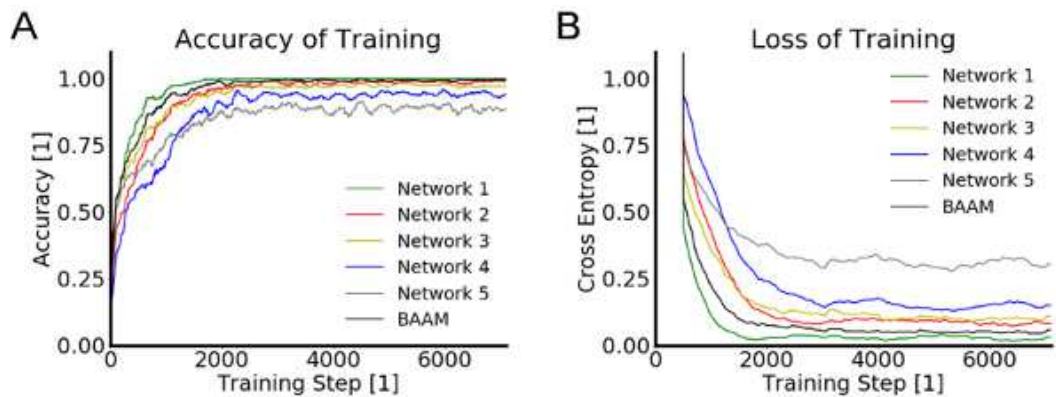}
\caption{Network performances during training. (A) Accuracy of age prediction for training set vs training iterations for $5$ sample networks and BAAM. (B) Smoothed loss value of age prediction for training set  vs training iterations for $5$ sample networks and BAAM.}
\label{fig:Appendix_Fig1}
\end{figure}

Various networks with different architectures and hyper parameters have been trained on the training set and applied on validation and test sets. The best performing network based on the accuracy of age prediction on  validation and test sets has been selected to analyze the application set. Although, there exist unlimited variations of these elements, only a subset could provide meaningful results for the aim of decoding the aging and rejuvenation processes. Therefore, we designed and tested different networks variations of network depth, Network layout (convolutional (C), pooling (P) and fully connected (F)) and the size of kernel in each convolutional layer. Here we present the comparison of network performances for five selected designed network in which the key parameters are tweaked to reach the highest prediction accuracy. The five networks have following details:
\begin{enumerate}
    \item Network 1:
    \begin{itemize}
        \item Depth: 7.
        \item Layout: C, C, P, C, C, P, F.
        \item Kernel sizes: 8, 16, 16, 32, 64, 64, 64.
        \item Number of iterations: 7000.
    \end{itemize}
    \item Network 2:
    \begin{itemize}
        \item Depth: 7.
        \item Layout: C, C, P, C, C, P, F.
        \item Kernel sizes: 16, 32, 32, 64, 128, 128, 128.
        \item Number of iterations: 7000.
    \end{itemize}
    \item Network 3:
    \begin{itemize}
        \item Depth: 10.
        \item Layout: C, C, P, C, C, P, C, C, P, F.
        \item Kernel sizes: 4, 4, 4, 8, 8, 8, 16, 32, 32, 32.
        \item Number of iterations: 7000.
    \end{itemize}
    \item Network 4:
    \begin{itemize}
        \item Depth: 8.
        \item Layout: C, C, P, C, C, P, F, F.
        \item Kernel sizes: 4, 8, 8, 16, 32, 32, 32, 32.
        \item Number of iterations: 7000.
    \end{itemize}
    \item Network 5:
    \begin{itemize}
        \item Depth: 4.
        \item Layout: C, C, P, F.
        \item Kernel sizes: 4, 8, 8, 8.
        \item Number of iterations: 7000.
    \end{itemize}
\end{enumerate}
Analyzing the accuracy and loss values of age prediction during training of different networks (Fig. \ref{fig:Appendix_Fig1}) shows that despite a correlation between depth of the its accuracy, going infinitely deeper does not necessary lead to better results. Evidently, the network with lowest number of layers (network 5) has the weakest performance by reaching $86\%$ accuracy with loss value of $0.27$ (Fig. \ref{fig:Appendix_Fig1}A-B gray line). While, BAAM, network 1, network 2 and network 3 with depth of $7$, $7$, $7$ and $10$ respectively have similarly the best performances ( accuracy:  Fig. \ref{fig:Appendix_Fig1}A-B black, green, red and yellow lines respectively). Therefore, after a certain depth, no extra performance is gained. The reached accuracy and loss values for all the networks after $7000$ training iterations is shown in Table \ref{tab:Appendix_1}.

\begin{table}[htp]
   \caption{Network performances. The accuracy and loss values of the networks with different architecture at the end of $7000$ training iterations.} 
   \label{tab:Appendix_1}
   \small % text size of table content
   \centering % center the table
   \begin{tabular}{SSSSSSSS} \toprule
        {Network} & {Depth} & {Validation Accuracy} & {Loss} & {Test Accuracy}\\
        \midrule
        1         &  7      & 0.99       & 0.00 & 5/8 \\  
        2         &  7      & 0.98       & 0.02 & 5/8 \\  
        3         &  10     & 0.98       & 0.11 & 6/8 \\ 
        4         &  8      & 0.92       & 0.18 & 5/8 \\  
        5         &  4      & 0.86       & 0.27 & 5/8 \\  
        BAAM      &  7      & 0.99       & 0.05 & 7/8 \\  \bottomrule
    \end{tabular}
\end{table}

\begin{figure}[htbp]
\centering
\includegraphics[width=\textwidth]{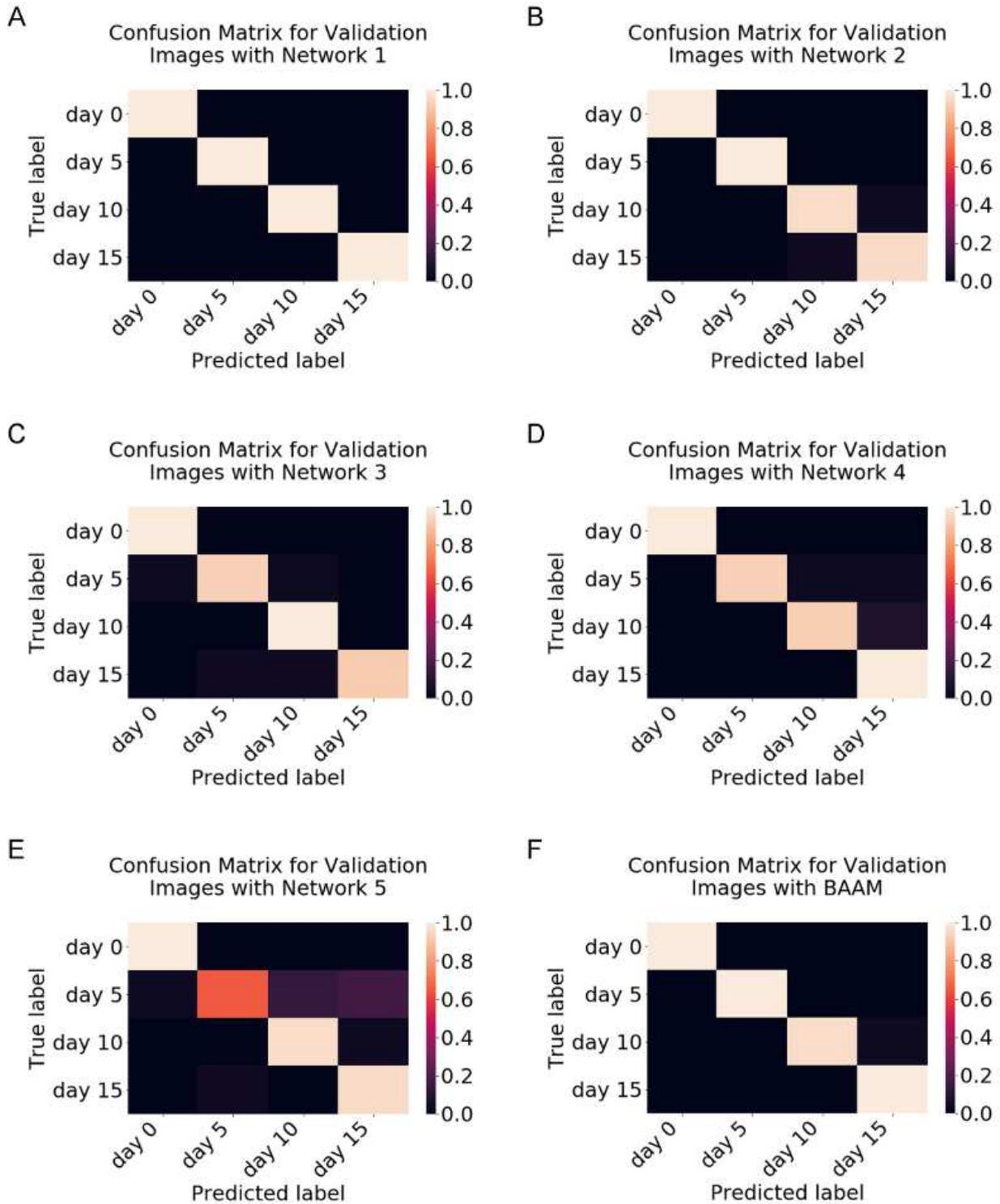}
\caption{Results of age prediction for validation set. (A)-(E): Networks 1-5 accordingly. (F): BAAM. In A-F Predicted age of samples are compared to their actual age. Values are normalized by the number of images per age class.}
\label{fig:Appendix_Fig2}
\end{figure}

Next, comparing the prediction results of different networks for validation set based on their confusion matrices further demonstrates that networks 1-4 (Fig. \ref{fig:Appendix_Fig2}A-D) have similar good performance and network 5 (Fig. \ref{fig:Appendix_Fig2}E) fails to predict day $5$ images correctly. It is further seen that BAAM (Fig. \ref{fig:Appendix_Fig2}F) provides the most diagonal confusion matrix hence is the best performing network. 

\begin{figure}[htbp]
\centering
\includegraphics[width=\textwidth]{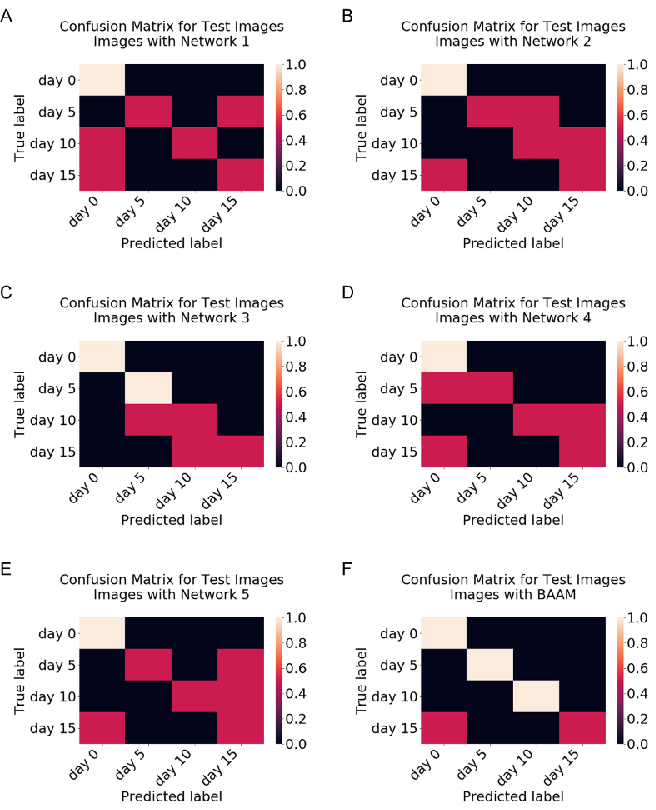}
\caption{Results of age prediction for test set. (A)-(E): Networks 1-5 accordingly. (F): BAAM. In A-F Predicted age of samples are compared to their actual age. Values are normalized by the number of images per age class.}
\label{fig:Appendix_Fig3}
\end{figure}

At last, the age prediction performance of the networks on the test set demonstrates the superiority of BAAM to other designed networks. While, networks $1$-$5$ achieve $5$, $5$, $6$, $5$ and $5$ correct predictions out of $8$ images of test set respectively (Fig. \ref{fig:Appendix_Fig3}A-E), BAAM reaches the highest number of correct age prediction with $7$ out $8$ (Fig. \ref{fig:Appendix_Fig3}F). 

The analysis of overall achievements of presented networks as examples of variance in characteristics of DNNs lead to designing the BAAM network . It over-performs the sample presented networks in age prediction for training, validation and test sets. Therefore, it is further utilized to decode the aging and rejuvenation processes.

\end{document}